\documentclass[10pt,twoside,spanish,english]{article}
\usepackage[T1]{fontenc}
\usepackage[latin9]{inputenc}
\usepackage[letterpaper]{geometry}
\usepackage{color}
\usepackage{babel}
\addto\shorthandsspanish{\spanishdeactivate{~<>.}}

\usepackage{float}
\usepackage{booktabs}
\usepackage{amsthm}
\usepackage{amsmath}
\usepackage{graphicx}
\usepackage{setspace}
\usepackage{esint}
\usepackage[unicode=true,pdfusetitle,
 bookmarks=true,bookmarksnumbered=true,bookmarksopen=true,bookmarksopenlevel=1,
 breaklinks=false,pdfborder={0 0 0},backref=false,colorlinks=false]
 {hyperref}
\usepackage{breakurl}

\makeatletter

\providecommand{\tabularnewline}{\\}

\numberwithin{equation}{section}
\numberwithin{figure}{section}
  \theoremstyle{remark}
  \newtheorem*{acknowledgement*}{\protect\acknowledgementname}

\usepackage{multicol}

\date{}

\makeatother

  \addto\captionsenglish{\renewcommand{\acknowledgementname}{Acknowledgement}}
  \addto\captionsspanish{\renewcommand{\acknowledgementname}{Agradecimientos}}
  \providecommand{\acknowledgementname}{Acknowledgement}

\begin{document}

\title{\noindent \textbf{A possible cosmological application of some thermodynamic
properties of the black body radiation in $n-$dimensional Euclidean
spaces.}}

\author{\textbf{Julian Gonzalez-Ayala, J. Perez-Oregon, Rub}\textsl{\'e}\textbf{n
Cordero and F. Angulo-Brown.}}

\maketitle
\begin{spacing}{0.10000000000000001}
\noindent \begin{center}
\textsl{Departamento de F\'isica, Escuela Superior de F\'isica y
Matem\'aticas, Instituto Polit\'ecnico Nacional, }\\
\textsl{Edificio 9, CP 07738, M\'exico D. F. }\\
\textsl{E-mail: julian@esfm.ipn.mx, jnnfr@esfm.ipn.mx, cordero@esfm.ipn.mx
and angulo@esfm.ipn.mx}\\

\par\end{center}
\end{spacing}
\begin{abstract}
In this work we present the generalization of some thermodynamic properties
of the black body radiation (BBR) towards an $n-$dimensional Euclidean
space. For this case the Planck function and the Stefan-Boltzmann
law have already been given by Landsberg and de Vos and some adjustments
by Menon and Agrawal. However, since then no much more has been done
on this subject and we believe there are some relevant aspects yet
to explore. In addition to the results previously found we calculate
the thermodynamic potentials, the efficiency of the Carnot engine,
the law for adiabatic processes and the heat capacity at constant
volume. There is a region at which an interesting behavior of the
thermodynamic potentials arise, maxima and minima appear for the $n-d$
BBR system at very high temperatures and low dimensionality, suggesting
a possible application to cosmology. Finally we propose that an optimality
criterion in a thermodynamic framework could have to do with the $3-d$
nature of the universe.
\end{abstract}
\begin{multicols}{2}
\newcommand{\dbar}{\mathchar'26\mkern-12mu d}

\section*{I. Introduction.}

Along much of the $20$th century there have been many contributions
to the development of a thermodynamic treatment of relativity and
cosmology \cite{key-1,key-2,key-3}. Since the extension of thermodynamics
to special relativity, first carried out by Planck \cite{key-4} and
by Einstein \cite{key-5}, and after to general relativity \cite{key-1},
until the thermodynamic approaches to cosmology \cite{key-2,key-3,key-6}.

On the other hand, Kaluza in 1921 \cite{key-7} and Klein in 1926
\cite{key-8} proposed a generalization of general relativity to unify
gravitation and electromagnetism by using a $5$-dimensional geometrical
model of space-time. Since then, many $n-$dimensional models within
cosmology have been proposed \cite{key-36,key-32,key-33,key-34,key-10,key-9,key-35,key-43,key-39}.
In 1989, Landsberg and De Vos \cite{key-11} made an $n$-dimensional
generalization of the Planck distribution, the Wien displacement law
and the Stefan-Boltzmann law for black body radiation (BBR) within
a zero curvature space. These authors also calculated the total internal
energy $U$ of radiation filling a hypervolume $V$ leading to the
recognition of the system as an ideal quantum gas obeying $pV=\varepsilon U$,
being $p$ the pressure and $\varepsilon$ a constant given by $\varepsilon=\frac{1}{n}$.
Recently, there has been a great interest in systems satisfying this
kind of equation of state with $\varepsilon$ positive or negative
for the case of an expanding accelerated universe \cite{key-12,key-13}.
Later, Menon and Agrawal \cite{key-14} modified the expression for
the Stefan-Boltzmann constant in $n$-dimensions found by Landsberg
and De Vos by using the appropriate spin-degeneracy factor of the
photon without affecting the curves of the normalized Planck spectrum
found by Landsberg and De Vos. In 1990 Barrow and Hawthorne investigated
the behavior of matter and radiation in thermal equilibrium in an
$n-$dimensional space in the early universe, in particular they calculated
the number of particles $N$, the pressure $p$ and the energy density
$u$ \cite{key-36}. In all these cases, the results have to do with
an isotropic and homogeneous average universe. Within this context,
in the present paper, in addition to the Landsberg-de Vos, and Barrow-Hawthorne
results for BB-radiation in an $n$-dimensional space, we calculate
for this system the thermodynamic potentials, the efficiency of a
Carnot cycle, the equation of adiabatic processes and the heat capacity
at constant volume. By using Planck units, it is possible to find
new and interesting information about the thermodynamic potentials.
A peculiar behavior in the potentials appear at very high temperatures
and low dimensionality, suggesting that a cosmological approach of
this phenomenon could give some clues about the nature of the dimensionality
of our space. The need of understanding this aspect of our universe
can be tracked to the ancient Greece \cite{key-46}. In modern times
this problem was first studied by Kant in 1746 \cite{key-44} and
later, among others by Ehrenfest in 1917 \cite{key-45}, Barrow in
1983 \cite{key-46}, Brandenberger and Vafa in 1989 \cite{key-6}
and Tegmark in 1997 \cite{key-47} (see also \cite{key-49,key-50,key-48,key-51}).
On the other hand, some corrections have been proposed in the context
of loop quantum gravity (LQG) by changing the dispersion relation
of photons in an $n-$dimensional flat space \cite{key-30}. Since
the region of interest is the one with very high temperature, we will
incorporate these corrections to the dispersion relation and the Planck
function in order to find the corresponding modification of the thermodynamic
potentials. The article is organized as follows: In Section II the
generalization of the thermodynamic potentials, the pressure and the
chemical potential to $n$ space dimensions is showed. In Section
III the equation for adiabatic processes is found. In Section IV the
efficiency of a Carnot cycle is calculated. In Section V an analysis
of critical points of the thermodynamic potentials is made, which
suggests a possible application to cosmology which is addressed in
Section VI. In Section VII a modification to the dispersion relation
stemming from LQG theories is incorporated into the analysis. Finally,
in Section VIII some conclusions are presented. Appendix A and B contain
some mathematical calculations.

\section*{II. Generalization to $n$ dimensions.}

For a black body in an $n$-dimensional space it is known \cite{key-36,key-11,key-14,key-15,key-16,key-26}
that the distribution of modes in the interval $\nu$ to $\nu+d\nu$
is,

\textcolor{black}{
\begin{equation}
g\left(\nu,n,V\right)d\nu=\frac{2\left(n-1\right)\pi^{\frac{n}{2}}V\nu^{n-1}}{\Gamma\left(\frac{n}{2}\right)c^{n}}d\nu,\label{eq:mod norm vib}
\end{equation}
}

where $V$ is the hypervolume, $n$ the dimensionality of the space
and $\Gamma\left(\frac{n}{2}\right)$ the gamma function evaluated
in $\frac{n}{2}$. The differential spectral energy is,

\begin{equation}
dU_{\nu}\left(\nu,T,V,n\right)=\frac{2\left(n-1\right)\pi^{\frac{n}{2}}Vh\nu^{n}}{\Gamma\left(\frac{n}{2}\right)c^{n}(e^{\frac{h\nu}{kT}}-1)}d\nu,\label{eq:dist esp energ Planck}
\end{equation}

which coincides with that in \cite{key-15} and \cite{key-16}. Integration
over the frequency gives the total internal energy,

\begin{equation}
U\left(T,n,V\right)=\frac{2\left(n-1\right)\pi^{\frac{n}{2}}V\left(kT\right)^{n+1}}{\Gamma\left(\frac{n}{2}\right)c^{n}h^{n}}\intop_{\chi=0}^{\infty}\frac{\chi^{n}}{e^{\chi}-1}d\chi,\label{eq:}
\end{equation}

where \textcolor{black}{$\chi\equiv\frac{h\nu}{kT}$.} The above integral
is equal to the product of the Riemann's Zeta function $\zeta\left(n+1\right)$
and $\Gamma\left(n+1\right)$, as it is shown in Appendix A. Then,
the energy volume density $u=U/V$ is,

\begin{equation}
u\left(T,n\right)=\frac{2\left(n-1\right)\pi^{\frac{n}{2}}\left(kT\right)^{n+1}\zeta\left(n+1\right)\Gamma\left(n+1\right)}{c^{n}h^{n}\Gamma\left(\frac{n}{2}\right)},\label{eq:energia vol}
\end{equation}

which reduces to the usual $T^{4}$-law for $n=3$. Eq. (\ref{eq:energia vol})
was found also by Barrow \cite{key-36}. The Helmholtz function\textcolor{black}{{}
$F(n,V,T)$ is given by the following expression \cite{key-17},}

\textcolor{black}{
\begin{equation}
F\left(n,V,T\right)=-kT\intop_{\nu=0}^{\infty}ln\left(Z_{ph}\left(\nu,T\right)\right)g\left(\nu,n,V\right)d\nu.\label{eq:helm}
\end{equation}
}

The last integral is solved in Appendix A. The final form of \textcolor{black}{$F(n,V,T)$
is,}

\[
F=-\frac{kT2\left(n-1\right)\pi^{\frac{n}{2}}V\left(kT\right)^{n}\varsigma\left(n+1\right)\Gamma\left(n+1\right)}{nc^{n}h^{n}\Gamma\left(\frac{n}{2}\right)}
\]

\textcolor{black}{
\begin{equation}
=-\frac{1}{n}U=-\frac{V}{n}u.\label{eq:helm int}
\end{equation}
}

Now, we calculate the entropy $S$. Since its relation with \textcolor{black}{$F(n,V,T)$
is given by \cite{key-17},}

\[
S=-\left(\frac{\partial F}{\partial T}\right)_{V}
\]

and according to Eq.(\ref{eq:helm int}), it turns out to be,

\noindent \textcolor{black}{
\begin{equation}
S=\left(\frac{n+1}{n}\right)\frac{U}{T}.\label{eq:entropia}
\end{equation}
}

From the Helmholtz function it is also possible to calculate the pressure
$p$ \cite{key-17},

\begin{equation}
p=-\left(\frac{\partial F}{\partial V}\right)_{T}.\label{eq:presion terminos de F}
\end{equation}

From Eq.(\ref{eq:helm int}) it immediately follows that,

\begin{equation}
p=\left(\frac{1}{n}\right)u\label{eq:presion}
\end{equation}

On the other hand, $U$, $F$ and $S$ are consistent with the Helmholtz
function expression \cite{key-17},

\begin{equation}
U=F+TS.\label{eq:-3}
\end{equation}

In the same manner, the enthalpy $H$ can be calculated as follows,

\begin{equation}
H\equiv U+pV=\left(\frac{n+1}{n}\right)U.\label{eq:Entalpia def}
\end{equation}

Comparing Eqs. (\ref{eq:entropia}) and (\ref{eq:Entalpia def}) we
see that $H=TS$. The Gibbs free energy is zero just as it happens
in the case $n=3$

\textcolor{black}{
\begin{equation}
G\equiv F+pV=0.\label{eq:gibbs}
\end{equation}
}

From the $3$-dimensional case it is known that the chemical potential
\textcolor{black}{$\mu$} of the photons is zero, and because this
is related to the Gibbs free energy by the equation \cite{key-17},

\textcolor{black}{
\[
\mu=\left(\frac{\partial G}{\partial N}\right)_{T,\: p},
\]
}

then it is concluded that this quantity is zero regardless the number
of space's dimensions. Since an isobaric process for the BBR is at
the same time isothermal, then the calorific capacity at constant
pressure $C_{p}\equiv\left(\frac{\partial H}{\partial T}\right)_{p}$
is undefined. On the other hand, $C_{V}$ can be calculated as usual, 

\begin{equation}
C_{V}=\left(\frac{\partial U}{\partial T}\right)_{V}=\frac{n+1}{T}U\label{eq:Cv explicita}
\end{equation}

This result agrees with the result obtained for the $3$-dimensional
case. All these results agree with the well-known 3-dimensional cases.
It is worthwhile to mention that the Planck's function has been obtained
within different contexts, from a formulation based in statistical
mechanics \cite{key-11,key-17}, considering special relativity \cite{key-18,key-19,key-20},
in the particle field theory framework and from electrodynamics. It
is equally valid for compactified and non-compactified dimensions,
and with exception of a multiplicative factor related with the spin
multiplicity of the photons, the same Plank's function is valid for
scalar particles, photons and gravitons \cite{key-36,key-11,key-12,key-15,key-16}.

\section*{III. Adiabatic Processes.}

From the first law of thermodynamics

\[
dU=\dbar Q-pdV,
\]

in an adiabatic process the exchange of heat is zero, then

\begin{equation}
dU=-pdV,\label{eq:cond adiabatica}
\end{equation}

because $U=uV$ and also $u=np$, then,

\[
nVdp+npdV=-pdV
\]

and

\begin{equation}
pV^{\frac{n+1}{n}}=constant.\quad\label{eq:adiabatica}
\end{equation}

which for $n=3$ agrees with the well known equation $pV^{\frac{4}{3}}=constant$
.

\section*{IV. Carnot cycle.}

By definition, the efficiency $\eta$ of a heat engine is given by

\[
\eta\equiv\frac{|W|}{|Q_{1}|}=\frac{|Q_{1}|-|Q_{2}|}{|Q_{1}|}=1-\frac{|Q_{2}|}{|Q_{1}|}.
\]

where $W$ is the work and $Q_{1}$ and $Q_{2}$ are the input and
output heats, respectively. Let's consider a heat engine working between
two reservoirs with temperatures $T_{1}$ and $T_{2}$ ($T_{1}>T_{2}$)
with black body radiation as the working substance. As we know, a
Carnot cycle consists of two isothermal processes at temperatures
$T_{1}$ and $T_{2}$ and two adiabatic processes. The calculation
for this efficiency is given in Appendix B where one obtains the already
familiar expression for the Carnot efficiency,

\[
\eta_{c}=1-\frac{T_{2}}{T_{1}}.
\]

Showing that the Carnot efficiency agrees with the $3$-dimensional
case. Moreover, this quantity is invariant under the number of dimensions.
What this means is that there is an upper limit to the efficiency
for a thermodynamic engine in any $n-$dimensional Euclidean space.
This is important since this fact establishes the second law of thermodynamics
as a more general principle \cite{key-52}.

\section*{V. Critical points of the thermodynamic potentials.}

Now we shall analyze the behavior of energy, entropy and Helmholtz
free energy densities (see Figs. 1 to 3). All the potentials have
a region of concavity or convexity. The units used are the so called
Planck's or natural units \cite{key-21}. In this scale the Planck's
temperature is $T_{P}=1.42\times10^{32}K$ and the energy of Planck
is $E_{P}=1.9561\times10^{9}J$. In each plot the most relevant region
is the one at high temperatures and low dimensions as can be seen
in Figures 1-3.

\begin{figure}[H]
\noindent \begin{centering}
\includegraphics[scale=0.8]{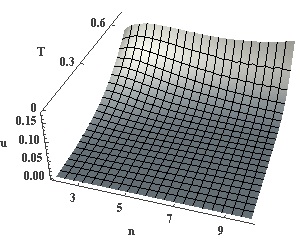}
\par\end{centering}

\protect\caption{Energy density $u\left(n,T\right)$ given by Eq. \ref{eq:energia vol}.}
\end{figure}

\begin{figure}[H]
\noindent \begin{centering}
\includegraphics[scale=0.8]{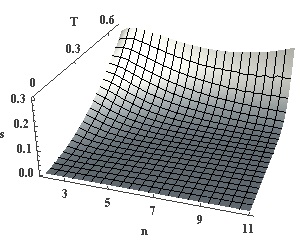}
\par\end{centering}

\protect\caption{Helmholtz free energy density $f\left(n,T\right)=F\left(V,n,T\right)/V$
(see Eq. \ref{eq:helm int}).}
\end{figure}

\begin{figure}[H]
\noindent \begin{centering}
\includegraphics[scale=0.8]{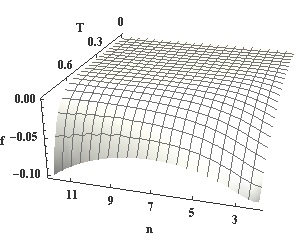}
\par\end{centering}

\protect\caption{Entropy density $s\left(n,T\right)=S\left(V,n,T\right)/V$ (see Eq.
\ref{eq:entropia}).}
\end{figure}

The functions $u\left(n,T\right)$, $f\left(n,T\right)$ and $s\left(n,T\right)$
do not have a global maximum nor minimum, but for a given temperature
we can find both local maxima and minima for each potential. These
critical points are located in a certain range of $n$ and $T$. Outside
this region the function presents a monotonic growth because of the
contribution of the term $\zeta\left(n+1\right)\Gamma\left(n+1\right)$.
One can analyze the first and second derivatives respect to the dimension
with the aim of determining whether the maxima and minima are restricted
to a certain dimensionality. For example, at constant temperature
one can goes from a minor to a major dimensionality, passing from
positive to negative derivatives (or vice versa). In this changes
of sign of the derivatives are the maxima (or minima) of the energy
density. As the temperature increases the minima and maxima get closer
to each other until they reach a saddle point (see Fig. 4). 

\begin{figure}[H]
\noindent \begin{centering}
\includegraphics{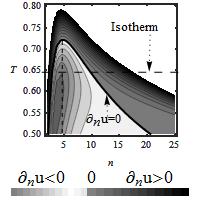}
\par\end{centering}

\protect\caption{First derivative in the dimension for the energy density $u\left(n,T\right)$
given by Eq. \ref{eq:energia vol}. The vertical dashed line marks
the location of the saddle point, located at $n<5$. The horizontal
dashed line represents an isothermal process, in which by increasing
the dimension from $n=1$ to $n=25$ the BBR system has an increasing
energy density $u$ (positive derivative), then it reaches a local
maximum and decreases (negative derivative) until it reaches a local
minimum and then $u$ increases again.}
\end{figure}

One point is of special interest, the point at which one cannot longer
obtain any of both extreme points (nor minima nor maxima). The saddle
point separates the regions of maxima an minima. This is summarized
in Table 1.

\begin{table}[H]
\noindent \begin{centering}
\begin{tabular}{ccccc}
\toprule 
 & $u\left(n,T\right)$ & $f\left(n,T\right)$ & $s\left(n,T\right)$ & $h\left(n,T\right)$\tabularnewline
\midrule
\midrule 
Maxima in & $n\leq4$ & $n>3$ & $n\leq4$ & $n\leq4$\tabularnewline
\midrule 
Minima in & $n>4$ & $n\leq3$ & $n>4$ & $n>4$\tabularnewline
\end{tabular}
\par\end{centering}

\protect\caption{Location of maxima and minima of energy, Helmholtz free energy, entropy
and enthalpy densities are restricted to certain dimensionalities.}
\end{table}

If a hypersphere is considered as the black body system, the total
potential can be found. The volume of this hypersphere of radius $R$
is given by the expression \cite{key-17},

\[
V=\frac{\pi^{\frac{n}{2}}}{\Gamma\left(\frac{1}{2}n+1\right)}R^{n}.
\]

In Figs. 5-7 the thermodynamic potentials $U\left(n,T.R\right)$,
$F\left(n,T,R\right)$, and $S\left(n,T,R\right)$ are shown (see
Eqs. \ref{eq:}, \ref{eq:helm int}, and \ref{eq:entropia} respectively),

\begin{figure}[H]
\noindent \begin{centering}
\includegraphics[scale=0.8]{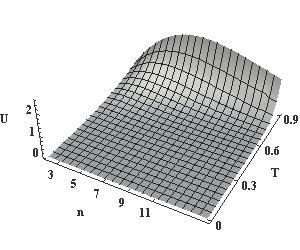}
\par\end{centering}

\protect\caption{Energy $U\left(n,T,R=.9\right)$ given by Eq. \ref{eq:}.}
\end{figure}

In these cases we cannot find any restriction in the dimensionality
at which the maxima or minima would appear. The variables $R$, $T$,
and $n$ can be combined in several ways in order to produce critical
points at any dimension.

\begin{figure}[H]
\noindent \begin{centering}
\includegraphics[scale=0.8]{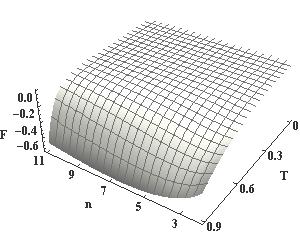}
\par\end{centering}

\protect\caption{Helmholtz free energy $F\left(n,T,R=.9\right)$ given by Eq. \ref{eq:helm int}.}
\end{figure}

\begin{figure}[H]
\noindent \begin{centering}
\includegraphics[scale=0.8]{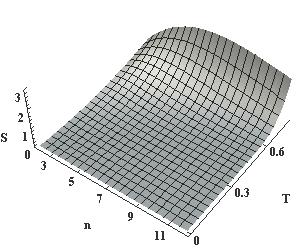}
\par\end{centering}

\protect\caption{Entropy $S\left(n,T,R=.9\right)$ given by Eq. \ref{eq:entropia}.}
\end{figure}

\section*{VIII. A possible simple application to cosmology.}

Since the works by Kaluza and Klein \cite{key-7,key-8}, many proposals
about universe's models with dimensionality different from three have
been published \cite{key-7,key-8,key-39,key-11,key-13,key-14,key-15,key-16,key-26,key-22,key-54,key-55}.
Remarkably this has been the case for papers stemming from the string,
d-branes and gauge theories \cite{key-32,key-33,key-34,key-35,key-13}.
However, nowadays we only have evidence for a universe with three
space and one time dimensions. In the present section we make a simple
thermodynamic analysis leading to a scenario of a universe with only
3 space dimensions. This scenario is present within the region of
high temperatures and low dimensionality mentioned in the previous
section.

Temperatures as high as $10^{32}K$ are only possible in a very primeval
time in the evolution of the universe. It is known that the period
dominated by radiation is from $t=10^{-44}s$ to $t\approx10^{10}s$
\cite{key-23}, that means, that in this period the universe could
have been well described by a spatially flat, radiation-only model
\cite{key-31}. Thus, considering the whole primeval universe as a
black body system in an Euclidean space is in principle a reasonable
approach.

In some textbooks \cite{key-24,key-25} it is referred that the critical
energy density at the time of $t\approx1t_{P}$ was about $u\approx1E_{P}/V_{P}$
($V_{P}$ is a Planck volume) and the temperature approximately $T\approx1T_{P}$
. The energy density and temperature obtained when the maximum of
the energy density is located at $n=3$ (see Fig 2) are surprisingly
near of the values above referred. 

According to string theories at some point (at the end of the Planck
epoch) the rest of the dimensions collapsed and only the known 3-dimensional
space grew bigger. The remaining question is: why do we live in a
$3-d$ universe?

Let's suppose that by some sort of mechanism the dimension of space
was reduced from a given initial value of the dimension to a final
state with $n=3$. In this scenario a change of energy can be obtained
as it is shown in Table 2. 

\begin{table}[H]
\begin{centering}
\begin{tabular}{ccc}
\toprule 
$n$ & $u\;\left[E_{p}/V_{P}\right]$ & $E$$\;[GeV/V_{P}]$\tabularnewline
\midrule
\midrule 
3 & 0.49 & 6.02E18\tabularnewline
\midrule 
4 & 0.66 & 8.04E18\tabularnewline
\midrule 
9 & 2.74 & 3.34E19\tabularnewline
\midrule 
25 & 15000 & 1.83E23\tabularnewline
\bottomrule
\end{tabular}
\par\end{centering}

\protect\caption{Energy (Eq. \ref{eq:energia vol}) at $T=0.93T_{p}$ within a Planck
volume. $T=0.93T_{p}$ is an arbitrary value close to $T_{P}$. }
\end{table}

If such transitions would occur between an initial and a final dimension,
a difference of energy could exist. In some cases the ``missing''
energy represents a significant part of the total initial energy,
as can be seen in Table 3.

\begin{table}[H]
\begin{centering}
\begin{tabular}{cccc}
\toprule 
$n_{initial}$ & $n_{final}$ & $\Delta E_{p}$ & $\%E_{missing}$\tabularnewline
\midrule
\midrule 
4 & 3 & 0.16 & 25.177\tabularnewline
\midrule 
9 & 3 & 2.24 & 82\tabularnewline
\midrule 
10 & 3 & 3.42 & 87.42\tabularnewline
\midrule 
25 & 3 & 14999.7 & 99.9967\tabularnewline
\bottomrule
\end{tabular}
\par\end{centering}

\protect\caption{Change of energy at the temperature of $T=0.93T_{p}$ in a Planck
volume.}
\end{table}

The energy difference between an initial and the final state is quite
big, in each case bigger than the total energy density for $n=3$,
and certainly must be taken into account whenever this model could
be applied in such processes. In that case, the questions of where
did that energy go? and if could it be in the extra compactified dimensions?
may be relevant. For example, for a $9+1$ space-time, the $6$ extra
spatial dimensions are supposed to be compactified, with a compactification
related to the Planck scale \cite{key-15}. One can look at those
numbers and think in a different way: the universe should have been
$1.34$ times more energetic to form a $4$ dimensional space, $5.6$
times more energetic for a $9$ dimensional space and $30476$ times
more energetic to form a $25$ dimensional space. Except for $n=4$
the difference between any scenario is considerable.

In the previous section we saw that the maxima and minima are restricted
to certain dimensionality, thus it could be a worthy effort to explore
whether some thermodynamic property could intervene in the construction
of a $3-$dimensional universe. 

It is possible from this generalization of the Planck function to
an Euclidean $n-$dimension space, to incorporate some corrections.
Given the range of high temperatures, a modification of the dispersion
relation found in LQG can be considered.

\section*{IX. LQG corrections}

There is a large number of people in the scientific community that
accept the possibility of the existence of extra dimensions beside
the $3+1$ known space-time dimensions. If one is willing to give
the black body radiation a cosmological context, it is necessary to
ask about the applicability of the above ideas in the regime of very
high energies and temperatures. Nowadays, it is clear the necessity
of having a theory that unifies the quantum world with the general
relativity in order to understand better the possible scenarios of
the high energy and the physics near the Planck epoch. One path to
do this is by the modification of general relativity, that is quantum
gravity, and this is the scope given by some important research programs.
In the context of brane theories it is possible to modify the geometrical
part of the Einstein equations to incorporate extra dimensions, which
have had certain success in explaining the dark energy problem and
the interpretations of astronomical data \cite{key-30}. On the other
hand, the extra dimensions are considered to have the size of the
Planck scale and remain in certain way hidden to our detectors \cite{key-32,key-33,key-34,key-35}.
In a variety of quantum gravity treatments, string theory, LQG, doubly
special relativity, models based in non commutative geometry, there
exists a common topic: the dispersion relation modification. Some
of these are related to address black hole thermodynamic problems.
Nozari and Anvari in Ref. \cite{key-30} proposed a modification of
the Planck law. This was made by using a modified dispersion relation
showed in \cite{key-30,key-28,key-29}, in which the square of the
energy-momentum vector is
\begin{equation}
\vec{p}^{2}\simeq E^{2}-\mu^{2}+\alpha L_{P}^{2}E^{4}+\alpha'L_{P}^{4}E^{6}+\mathcal{O}\left(L_{P}^{6}E^{8}\right),\label{eq:p^2 modificada}
\end{equation}
where $L_{P}$ is the Planck length and it depends on the dimensionality
of the space-time \cite{key-29}, $E$ is the energy and $\alpha$
and $\alpha'$ take different values depending on the details of the
quantum gravity candidates. Finally the modified spectral energy density
showed in \cite{key-30} is,

\selectlanguage{spanish}%
\[
g\left(\nu\right)\nu d\nu=\frac{2\pi^{\frac{n}{2}}\left(n-1\right)\nu^{n}}{\Gamma\left(\frac{n}{2}\right)\left(e^{\nu/T}-1\right)}\{1+\left(\frac{3}{2}+\frac{n-1}{2}\right)\alpha\nu^{2}
\]

\selectlanguage{english}%
\begin{equation}
+\left[\alpha^{2}\left(-\frac{5}{8}+\frac{3\left(n-1\right)}{4}\right)+\alpha'\left(\frac{5}{2}+\frac{n-1}{2}\right)\right]\nu^{4}\}\label{eq:Densidad Planck Modificada}
\end{equation}

In this way, the internal energy density is given by the following
expression (see \cite{key-30}),

\selectlanguage{spanish}%
\[
u\left(n,T\right)=\frac{2\pi^{\frac{n}{2}}\left(n-1\right)T^{n+1}}{\Gamma\left(\frac{n}{2}\right)}\{\Gamma\left(n+1\right)\zeta\left(n+1\right)+
\]

\[
\left(\frac{3}{2}+\frac{n-1}{2}\right)\alpha T^{2}\Gamma\left(n+3\right)\zeta\left(n+3\right)+
\]
\[
\left[\alpha^{2}\left(-\frac{5}{8}+\frac{3\left(n-1\right)}{4}\right)+\alpha'\left(\frac{5}{2}+\frac{n-1}{2}\right)\right]\times
\]

\begin{equation}
T^{4}\Gamma\left(n+5\right)\zeta\left(n+5\right)\}.\label{eq:u modificada grav cu=0000E1ntica}
\end{equation}

\selectlanguage{english}%
and from Eq. (\ref{eq:helm}) and Appendix A, the Helmholtz free energy
density now is,

\[
f\left(n,T\right)=-\frac{2\pi^{\frac{n}{2}}\left(n-1\right)T^{n+1}}{\Gamma\left(\frac{n}{2}\right)}\{\frac{\Gamma\left(n+1\right)\zeta\left(n+1\right)}{n}+
\]
\[
\frac{\left(\frac{3}{2}+\frac{n-1}{2}\right)\alpha T^{2}\Gamma\left(n+3\right)\zeta\left(n+3\right)}{n+2}
\]
\[
+\left[\alpha^{2}\left(-\frac{5}{8}+\frac{3\left(n-1\right)}{4}\right)+\alpha'\left(\frac{5}{2}+\frac{n-1}{2}\right)\right]\times
\]
\begin{equation}
\frac{T^{4}\Gamma\left(n+5\right)\zeta\left(n+5\right)}{n+4}\}.\label{eq:f LQG mod}
\end{equation}

\begin{figure}[H]
\noindent \begin{centering}
\includegraphics{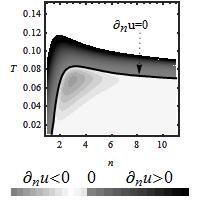}
\par\end{centering}

\protect\caption{First derivative in the dimension for the energy density $u\left(n,T\right)$
given by Eq. \ref{eq:u modificada grav cu=0000E1ntica}. The values
of $\alpha=0.1$ and $\alpha'=2.28$ are such that the curves showed
in \cite{key-30} are reproduced. }

\end{figure}

One relevant aspect is that it is possible to get the same interesting
behavior of the thermodynamic potential densities mentioned before,
but one order of magnitude smaller in the temperature than in the
model with no LQG modifications, as can be seen in Fig. 8 (see Fig.
4 for comparison).

\section*{X Conclusions}

It has been presented the generalizations of the entropy, free energies
of Helmholtz and Gibbs, enthalpy, pressure, the equation for adiabatic
processes, the Carnot cycle efficiency, and the chemical potential
for the black body radiation in an $n$-dimensional Euclidean space. 

This flat space generalization with reasonable suppositions could
be a good approximation to model the very early universe. The peculiar
behavior of the thermodynamic potential densities in the region of
low dimensionality-high temperature could be profoundly linked with
the 3-dimensional character of our space since the beginning of time.
In particular, the maximization of some thermodynamic functions seems
a suitable scenario to choose $n=3$ since the very early universe.
Finally, a correction that takes into account modifications to the
dispersion relations stemming from loop quantum gravity are incorporated
into the analysis.

\section*{Appendix A}

\subsection*{{\small{}Riemann's zeta function
\[
\zeta\left(x\right)=\frac{1}{1^{x}}+\frac{1}{2^{x}}+\frac{1}{3^{x}}+...
\]
\begin{equation}
\zeta\left(x\right)=\frac{1}{\Gamma\left(x\right)}\intop_{0}^{\infty}\frac{u^{x-1}}{e^{u}-1}du\;\;\;\;\;\;\; x>1.\label{eq:zeta Riemann}
\end{equation}
}}

\subsection*{{\small{}Integrating $\quad\intop_{\chi=0}^{\infty}\chi^{n-1}ln\left(1-e^{-\chi}\right)d\chi$
by parts.}}

{\small{}
\[
u=ln\left(1-e^{-\chi}\right),\;\;\; du=\frac{e^{-\chi}}{1-e^{-\chi}}d\chi=\frac{1}{e^{\chi}-1}d\chi
\]
}{\small \par}

{\small{}
\[
dv=\frac{1}{n}d\chi^{n},\;\;\;\;\;\;\;\;\;\;\;\; v=\frac{\chi^{n}}{n}\;\;\;\;\;\;\;\;\;\;\;\;\;\;\;\;\;\;\;\;\;\;\;\;
\]
\[
\intop_{\chi=0}^{\infty}\chi^{n-1}ln\left(1-e^{-\chi}\right)d\chi=\frac{\chi^{n}}{n}ln\left(1-e^{-\chi}\right)\mid_{\chi=0}^{\infty}
\]
\[
-\frac{1}{n}\intop_{\chi=0}^{\infty}\frac{\chi^{n}}{e^{\chi}-1}d\chi.
\]
}{\small \par}

{\small{}The first member of the right side of the equation, by L'Hopital's
rule is zero 
\[
\underset{\chi\rightarrow0}{lim}\frac{ln\left(1-e^{-\chi}\right)}{\chi^{-n}}=\underset{\chi\rightarrow0}{lim}-\frac{\chi^{n+1}}{n\left(e^{\chi}-1\right)}
\]
\[
=\underset{\chi\rightarrow0}{lim}-\frac{\left(n+1\right)\chi^{n}}{ne^{\chi}}=0,
\]
\[
\underset{\chi\rightarrow\infty}{lim}\frac{ln\left(1-e^{-\chi}\right)}{\chi^{-n}}=\underset{\chi\rightarrow\infty}{lim}-\frac{\chi^{n+1}}{n\left(e^{\chi}-1\right)}
\]
\[
=\underset{\chi\rightarrow\infty}{lim}-\frac{\left(n+1\right)\chi^{n}}{ne^{\chi}}=...=\underset{\chi\rightarrow\infty}{lim}-\frac{\left(n+1\right)!}{n^{2}e^{\chi}}=0.
\]
}{\small \par}

{\small{}And because of Eq.(\ref{eq:zeta Riemann}) 
\[
\intop_{\chi=0}^{\infty}\chi^{n-1}ln\left(1-e^{-\chi}\right)d\chi=-\frac{1}{n}\zeta\left(n+1\right)\Gamma\left(n+1\right).
\]
}{\small \par}

\section*{Appendix B. Carnot efficiency{\small{} }}

From the fist law of thermodynamics
\[
dQ=dU+pdV=npdV+nVdp+pdV
\]
because $p=p\left(T\right),$in an isothermal path $p=const$ and
$dp=0$,
\[
dQ=(n+1)pdV
\]
\begin{equation}
Q_{1}=(n+1)\intop_{1}^{2}p_{1}dV=(n+1)p_{1}\intop_{1}^{2}dV=(n+1)p_{1}(V_{2}-V_{1})\label{eq:calor1}
\end{equation}
and
\begin{equation}
Q_{2}=(n+1)p_{3}\intop_{3}^{4}dV=(n+1)p_{3}(V_{4}-V_{3})\label{eq:calor 2}
\end{equation}
\begin{equation}
\Rightarrow\eta=1-\frac{p_{3}(V_{3}-V_{4})}{p_{1}(V_{2}-V_{1})},\label{eq:eta1}
\end{equation}
{\small{}
\[
\eta=1-\frac{p_{3}(V_{3}-V_{4})}{p_{1}(V_{2}-V_{1})},
\]
from Eq. (\ref{eq:adiabatica})
\[
p_{1}V_{1}^{\frac{n+1}{n}}=p_{3}V_{4}^{\frac{n+1}{n}}\qquad y\qquad p_{1}V_{2}^{\frac{n+1}{n}}=p_{3}V_{3}^{\frac{n+1}{n}},
\]
then
\[
V_{1}p_{1}^{\frac{n}{n+1}}=V_{4}p_{3}^{\frac{n}{n+1}}\qquad y\qquad V_{2}p_{1}^{\frac{n}{n+1}}=V_{3}p_{3}^{\frac{n}{n+1}},
\]
\[
p_{1}^{\frac{n}{n+1}}(V_{2}-V_{1})=p_{3}^{\frac{n}{n+1}}(V_{3}-V_{4}),
\]
\[
\frac{(V_{3}-V_{4})}{(V_{2}-V_{1})}=\frac{p_{1}^{\frac{n}{n+1}}}{p_{3}^{\frac{n}{n+1}}},
\]
taking a look to Eq. (\ref{eq:eta1}) and because $p_{2}=p_{3}$,
\[
\eta_{c}=1-\frac{p_{1}^{\frac{n}{n+1}}p_{2}}{p_{2}^{\frac{n}{n+1}}p_{1}}=1-\frac{p_{2}^{\frac{1}{n+1}}}{p_{1}^{\frac{1}{n+1}}}
\]
we know that,
\[
p=a_{n}T^{n+1}
\]
\[
\Rightarrow p^{\frac{1}{n+1}}=\alpha T,
\]
finally arriving to the desired expression for the Carnot efficiency,
\[
\eta_{c}=1-\frac{T_{2}}{T_{1}}.
\]
}{\small \par}
\begin{acknowledgement*}
We want to thank partial support from COFAA-SIP-EDI-IPN and SNI-CONACYT,
M\textsl{\'E}XICO.
\end{acknowledgement*}

\end{multicols}{}
\end{document}